\documentclass[preprint]{aastex}
\include{epsf}
\begin{document}

\title{HST/WFPC2 proper motions in two bulge fields:
kinematics and stellar population of the Galactic bulge\altaffilmark{1}}
\altaffiltext{1}{Based on observations made with the NASA/ESA Hubble
Space Telescope, obtained
at the Space Telescope Science Insitute, which is operated by the
Association of Universities for Research in Astronomy, Inc., under
NASA contract NAS 5-2655.  These observations are associated with
proposal GO-8250.}

\author{Konrad Kuijken}
\affil{Kapteyn Institute, PO Box 800, 9700 AV Groningen, The Netherlands
\altaffilmark{2}}
\altaffiltext{2}{Visiting scientist, Dept. of Theoretical Physics,
Univ. of the Basque Country}
\email{kuijken@astro.rug.nl}

\and

\author{R. Michael Rich}
\affil{Division of Astronomy, Department of Physics and Astronomy, UCLA,
LA  CA  90095-1562}
\email{rmr@astro.ucla.edu}

\begin{abstract}

We report proper motion dispersions for stars in the direction of
two fields of the Galactic bulge, using HST/WFPC2 images taken six
years apart.  Our two fields  are Baade's Window
$(l,b)=(1.13^\circ,-3.77^\circ)$ and Sgr I $(l,b)=(1.25^\circ ,-2.65^\circ)$.
Our proper motion dispersions are in good agreement with prior ground-
and space-based proper motion studies in bulge fields, but in contrast
to some prior studies, we do not exclude any subset of stars from
our studies.  In Baade's Window, we find the $l$ and $b$ proper motion
dispersions are 2.9 and 2.5 mas/yr, while in Sgr I, they are 3.3 and
2.7 mas/yr, respectively.  For the first time, we can clearly
separate the foreground disk stars out from the bulge because of their
large mean apparent proper motion.  The population with non-disk kinematics (which we conclude to be the bulge) has an old main sequence
turnoff point, similar to those found in old,
metal rich bulge globular clusters while those stars selected to
have disk kinematics lie on a fully populated main sequence.
Separating main sequence stars
by luminosity, we find strong evidence that the bulge population is
rotating, largely explaining observations of proper motion anisotropy
in bulge fields.  Because we have isolated such a pure sample of
stars in the bulge, we have one of the clearest demonstrations that
the old stellar population of the inner bulge/bar is in fact rotating.

\end{abstract}

\keywords{Galaxy: bulge, kinematics and dynamics, stellar conent;
methods: data analysis}

\section{Introduction}

The bulge of our Galaxy is interesting for a number of reasons. It is
the nearest galactic bulge to the sun, and represents a unique place
to study the stellar populations and stellar dynamics of such objects
in detail. Such analysis can provide important information for our
understanding of how bulges formed their stellar populations, what
gravitational potential they sit in, and how they came to have the
structure they do.  In the case of our Galactic bulge, stellar motions
are of great interest both in testing the hypothesis that our bulge
is in fact a bar (e.g.\ \citealt{zhaoetal96}) and in the
modeling of microlenses, the overwhelming majority of which are
seen in the direction of the bulge \citep{alcock00}.

There are several complications in trying to build an understanding of
our bulge. Close to the galactic plane, extinction by foreground dust
is large and uneven (even on very small scales) and effectively
limits optical work to a few ``windows'' of low and relatively uniform
extinction.  More troubling is the difficulty that the stellar population
seen in these directions samples everything along the line of sight.
Simple exponential models for the
disk, for example, predict that in Baade's Window at least half of the
stars visible are actually disk stars, and not bulge stars.
Employing photometry alone, it is not possible to effectively
sort the populations; bulge and
disk populations overlap in color, especially near the turnoff
\citep{holtz98} greatly complicating the use of HST-derived
CMDs for age determination.  Blue stragglers extending brighter than the
turnoff in an old population overlap with the main sequence locus
of a young population.

Among the first stellar populations imaged in 1994 with the repaired
WPFC2 on board HST were fields near two of the low extinction regions
originally identified by Walter Baade: what is now known as Baade's
Window $(l,b)=(1.13^\circ,-3.77^\circ)$,  and Sgr I $(l,b)=(1.25^\circ
,-2.65^\circ)$.   A discussion of the photometry
and luminosity function of the Baade's Window field is given in
\citet{holtz98}.

We noticed that the fields were ripe for a revisit
with the aim of measuring proper motions and we proposed successfully
(GO-8250).  Although HST has been used to measure proper motions of
field stars in the rough vicinity of the Galactic Center, these two
fields are of special interest because they lie well within the {\sl
COBE} bulge \citep{dwek95} and, in the case of Baade's Window,
abundances are measured \citep{mcw94} and many other
studies have been done.  Combining proper motions with precision
photometry might make it possible to separate the observed populations
based on their kinematics.  The only previous study of bulge proper
motions in Baade's Window was based on photographic plates
\citep{spaen92} and reported motions only of
stars thought to be candidate red giants in the bulge, by \citet{arp65}.
By systematically excluding the bluer disk stars, and by only
measuring a few hundred of the brightest giants, this study, while
pioneering, leaves much of the problem ripe for inquiry.  Hence our
decision to obtain second epoch images of both bulge fields, using
WFPC2 on board HST.

\section{Observations and proper motion measurements}

In August of 2000 we obtained HST/WFPC2 observations of two bulge
fields for which archival images suitable as first-epoch observations
existed. The coordinates and observations are listed in
Table~\ref{tab:obs}.

\begin{table}
\caption{HST/WFPC2 data used in this paper}
\begin{tabular}{lllll}
\tableline\tableline
Field & Epoch & Exp (s)  & Remarks& $\alpha,\delta\ (2000)$\\
\tableline
Baade Window & 13 Aug 1994 & 20, 200 ($\times2$), 1000 ($\times2$) & undithered& 18 03 10\\
 \ \ $(l=1.13,\ b=-3.77)$ &2 Sep 1995  & 40 ($\times2$), 200 ($\times4$), 400 ($\times3$) & undithered &  $-29$ 51 45\\
 & 30 Aug 2000 & 40 ($\times8$), 400 ($\times8$) & 4-way dithered &\\
\hline
Sgr-I field & 21 Aug 1994 & 10, 100, 1000 ($\times3$) & undithered& 17 59 00.5\\
 \ \ $(l=1.25,\ b=-2.65 )$  & 8 Aug 2000 & 40 ($\times8$), 400 ($\times8$) & 4-way dithered& $-29$ 12 14\\
\tableline
\end{tabular}
{ Observations in 2000 obtained as part of this
programme used filter F814W; of the archival 1994 and 1995 data sets,
images obtained with the F814W were used for the proper motion
determinations, and additionally F555W images were used for the
color-magnitude diagrams.}
\label{tab:obs}
\end{table}

The archival data, though they consist of multiple exposures, are
not dithered. This compromises our ability to centroid stars
accurately, as the images on the wide-field camera CCDs are very much
undersampled. We therefore chose to observe the second epoch data
(henceforth referred to as Y2K data) in the F814W filter, which gives
the widest, and hence least undersampled, point-spread function
(PSF). We also decided to dither the Y2K exposures in a regular 2x2
pattern of offsets (with a step of 0.55 arcsec, or 5.5 pixels on the
WF CCDs), ensuring that the uncertainties in the proper motions would
not be dominated by the newer epoch.

It is easy to show that the 1-$\sigma$ uncertainty in the centroid of
a stellar image of full-width at half maximum FWHM that is detected at
significance (S/N) is given by
\[
0.7 \times \hbox{FWHM} / \hbox{(S/N)}
\]
where the numerical coefficient has a constant value for
Moffat-profile PSF's with asymptotic power-law slopes ranging from 2
to infinity, the latter corresponding to a gaussian PSF (see
Appendix).
With HST, therefore (FWHM$\simeq0.12$arcsec in
F814W), a star detected at 25-$\sigma$ could be centered to about
3.4mas (1/30th of a pixel). Over a 7-year baseline, proper motions can
be measured for such stars to an accuracy of 0.7mas/yr. At the
distance to the bulge of about 8kpc this corresponds to about 25km/s.

\citet{ak00} and \citet{ibata98} describe ways in which astrometry can
be performed in dense star fields using multiple dither
observations. Essentially, these methods simultaneously derive a PSF
(convolved with the pixel shape) that is consistent with the multiple
dither observations of the same star field, and centroids for all
stars in the field.

In the case of undersampled data, such as our first-epoch data, it is
necessary to have independent knowledge of the shape of the PSF in
order to be able to determine an accurate centroid. In dense star
fields this information can be extracted provided the undersampling is
not too severe: the stars will be positioned randomly with respect to
the pixel grid of the detector, and so the ensemble of stars does in
fact sample the PSF at a wide range of phases. Our method consists in
a modification of the \citet{ak00} approach. For each exposure we
start from an analytic PSF model, and determine best-fit centers and
intensities for the 100 brightest unsaturated stars in the frame that
are not affected by neighbours. These centers and intensities are then
used to build an ensemble-averaged PSF from these stars. An analytic
fit of somewhat higher order is then made to this new sampling of the
PSF, with a robust outlier rejection. This new PSF is then used to
define new centers and intensities, etc. This process iterates to
convergence quite rapidly. In practice we use a gaussian PSF
multiplied by polynomials as our model.

Armed with a PSF for each observation, we determine the proper motions
as follows. First we combined all undithered data from the 1995
epochs, filter by filter, and identify 20-$\sigma$ detections in the
combined image using the DAOFIND package. This master coordinate list
for each CCD is then transformed using bright reference stars to align
it with the positions in each individual exposure. For each F814W
exposure a PSF fit is then performed near each transformed master list
position, and the best-fit center and intensity recorded. The result
is a set of up to 30 (21) position measurements for each star in the
Baade Window (Sgr-I) fields, referred to the pixel positions of the
different exposures, and spread over three (two) epochs.

\begin{figure}
\epsfxsize=\hsize\epsfbox{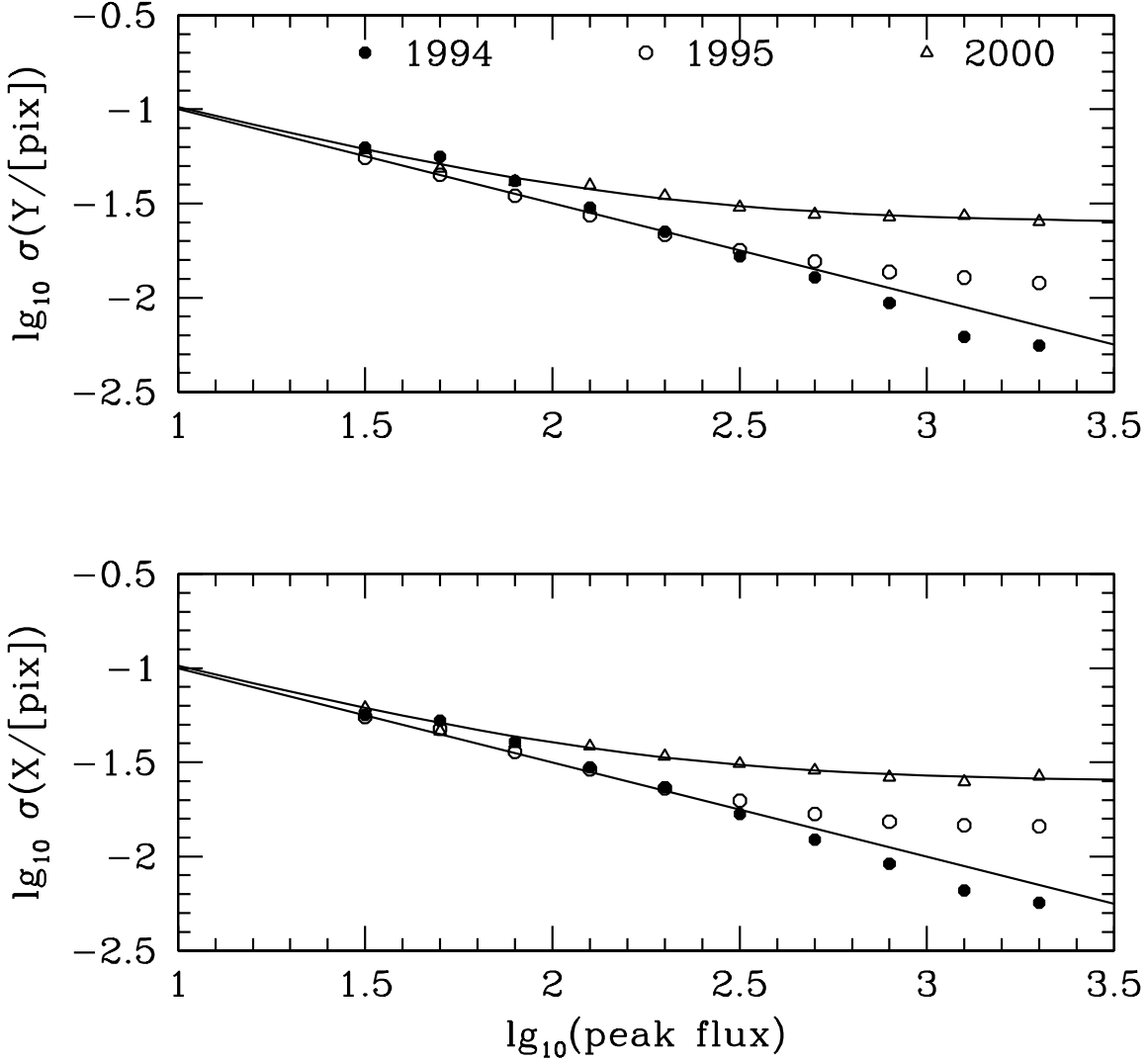}
\caption{The rms residuals in $x$ (top) and $y$ (bottom) pixel position
from the proper motion fits for the WF2 images of Baade's Window. The
plot illustrates that for the fainter stars the dominant source of
error is photon noise, but that bright stars suffer from some residual
systematic errors which are not removed by our procedure. This
manifests itself as a higher residual error in the properly dithered,
epoch 2000 data. The 1994 and 1995 data, which were not dithered, do
not show this extra residual, indicating that any such systematics
affecting those data have been absorbed into the proper motion
fit. The solid lines represent a theoretical fit of the effect of
photon statistics alone (lower line) and of an additional constant
systematic error (upper curve). The upper curve has been used to
generate error estimates on all positions and proper motions.
A similar plot for PC data does not show such an additional systematic
error.}
\end{figure}

These position measurements contain the proper motion information we
seek, but also plate solution mappings from one exposure to another,
and residual systematic centroiding errors. An iterative procedure,
consisting of the following loop, separates the various effects: (i)
find an average transformation that maps the star positions of each
exposure onto those of a reference exposure, and apply this
transformation to the positions; (ii) determine proper motions for
each star based on these transformed positions, and determine position
residuals for each position measurement from the best-fit proper
motion; (iii) determine the average position residual as a function of
pixel phase of each stellar image, and subtract this residual from
each measurement; (iv) determine the average position residual as a
function of phase with respect to the `34th-row' effect \citep{ak99}
and subtract this. At each pass through the loop a new proper motion
solution is made for each star, using a weighted linear least-squares
fit with rejection of outliers. The weights are deduced from the
estimated centroiding error of each image on the basis of the number
of photons detected, including an estimated residual systematic error
(see Figure 1).

Effects (iii) and (iv) are only detectable with dithered data, and so
we have had to assume that the systematic rediduals with pixel phase
and 34th-row phase (both on the order of 2--5\% of a pixel) apply also
to the earlier epochs. Any error we make in this assumption is not
easily detectable in our data: any such positional error in the first
epoch will simply be absorbed into the proper motion measurement.
Two pieces of evidence give us confidence that our results are not
seriously affected by such residual systematics:
\begin{enumerate}
\item
The multiple dithers observed in our Y2K data provide a test of the
procedure. With many dithered observations at the same epoch, residual
errors show up as extra scatter about the best-fit position that cannot
be absorbed into the proper-motion solution.
Figure 1 shows that for the
fainter stars our errors are safely dominated by photon statistics,
and that systematic errors set in at the 0.03-pixel level on the
undersampled WF detectors (corresponding to 20km/s over a 7-year
baseline at the distance of the bulge).
\item
Comparisons of the distributions of the proper motions
between the PC and WF detectors shown below do not indicate that our
WF proper motions contain larger errors than those derived from the
PC.
\end{enumerate}

Using this technique, we obtained proper motion measurements for some
20,000 stars in the Sgr-I field, and for 15,000 stars in Baade's
window. In addition, we used the 1994/1995 archival data for each field to
derive magnitudes in the F814W and F555W filters, using the magnitude
calibrations given in \citet{holtz95}.

Note that the proper motions derived are {\em relative}: we
arbitrarily assign the mean proper motions of the complete sample of
stars in each field to be zero in both components. Absolute (i.e.,
with respect to an inertial frame) proper motion measurements would
require extragalactic sources such as qso's to be identified in these
fields. Perhaps future spectral or multi-color studies can provide
this extragalactic reference frame.

\section{Analysis}

\begin{figure}
\epsfxsize=0.45\hsize\epsfbox{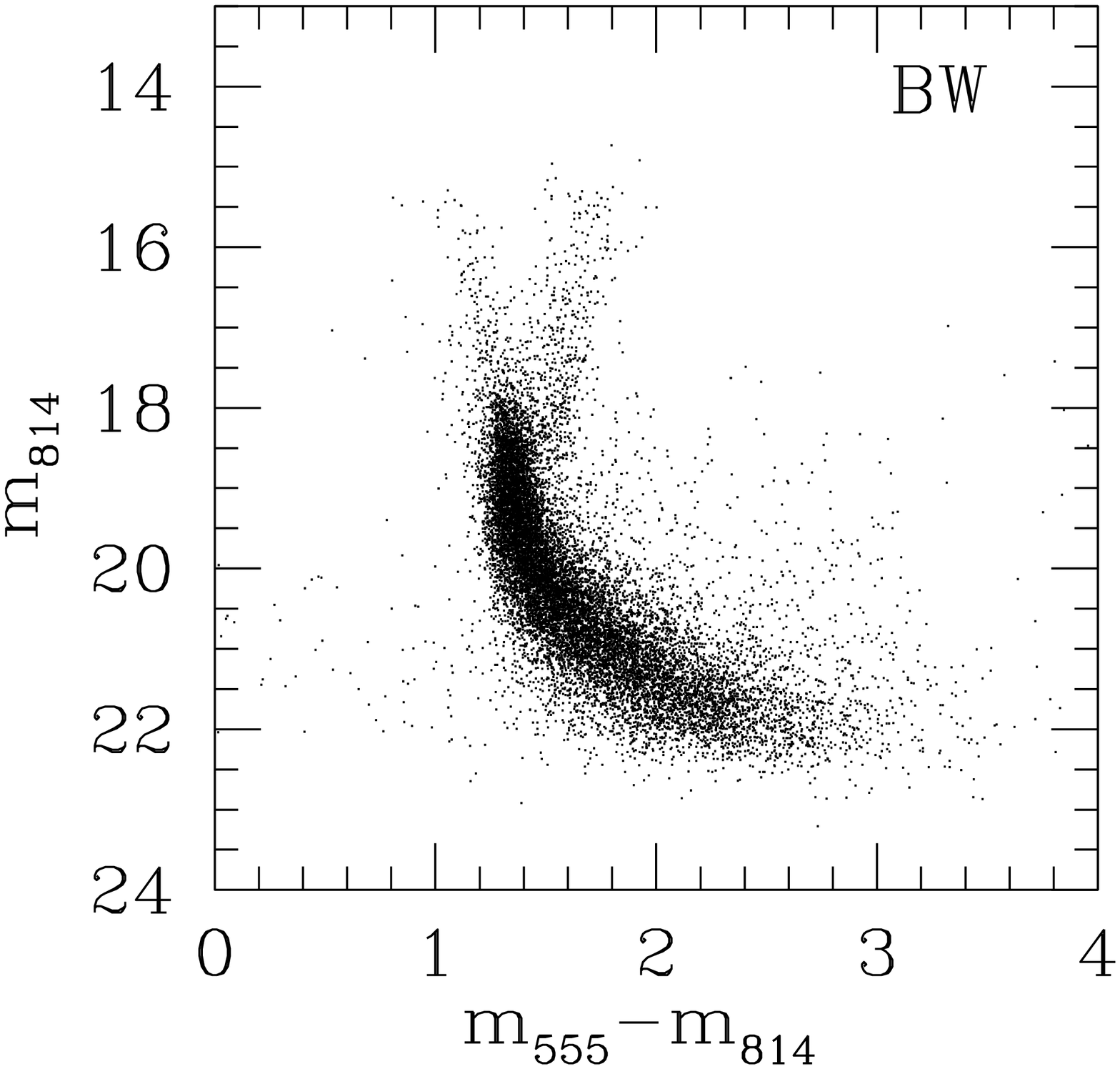}
\epsfxsize=0.45\hsize\epsfbox{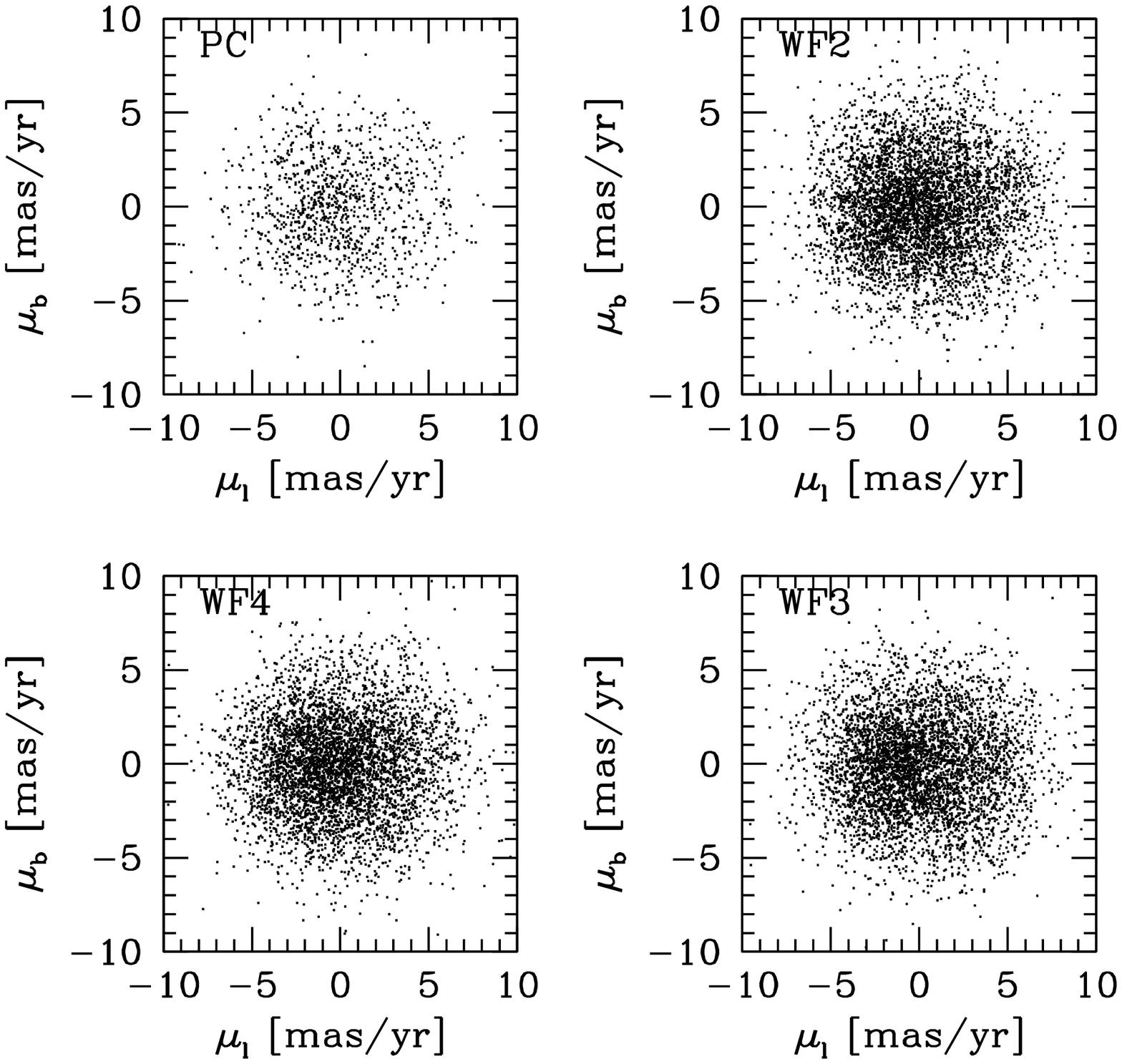}
\caption{F814W/F555W color-magnitude diagram for the Baade Window
field, and proper motion distributions for the individual CCD frames of the
WF and PC.}
\end{figure}

\begin{figure}
\epsfxsize=0.45\hsize\epsfbox{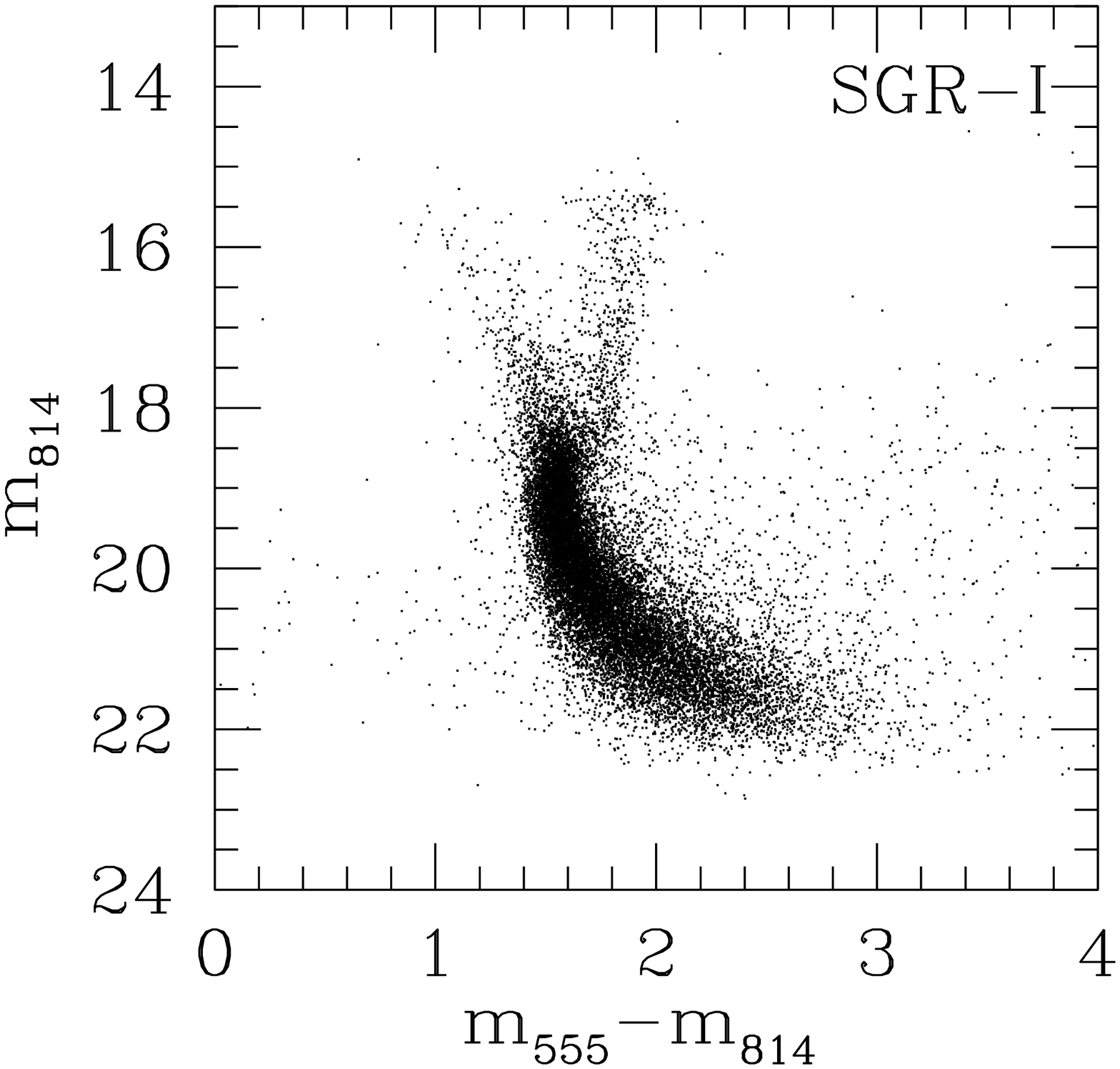}
\epsfxsize=0.45\hsize\epsfbox{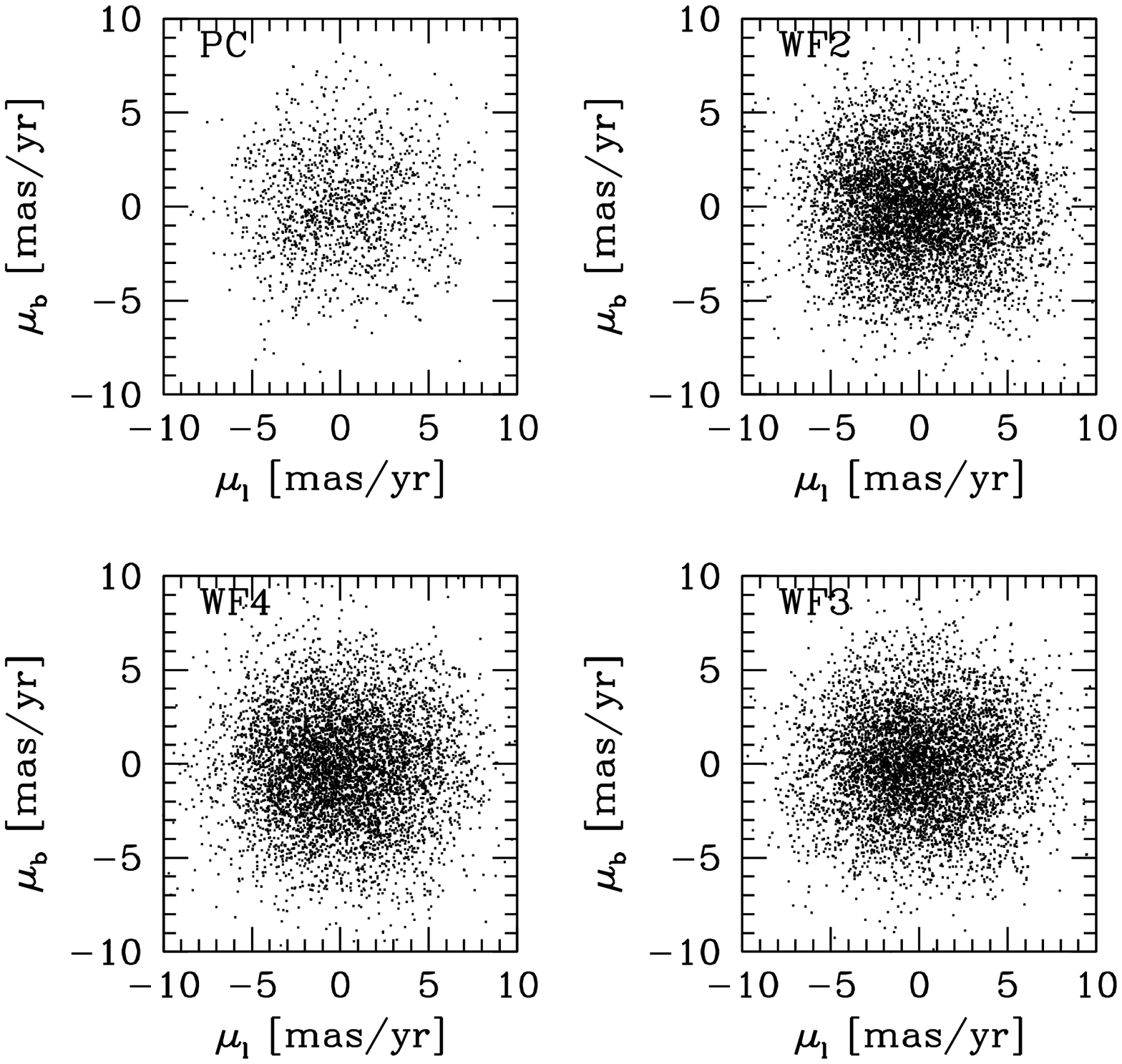}
\caption{F814W/F555W color-magnitude diagram for the SGR-I field, and
proper motion distributions for the individual CCD frames of the WF
and PC.}
\end{figure}

\begin{figure}
\epsfxsize=\hsize\epsfbox{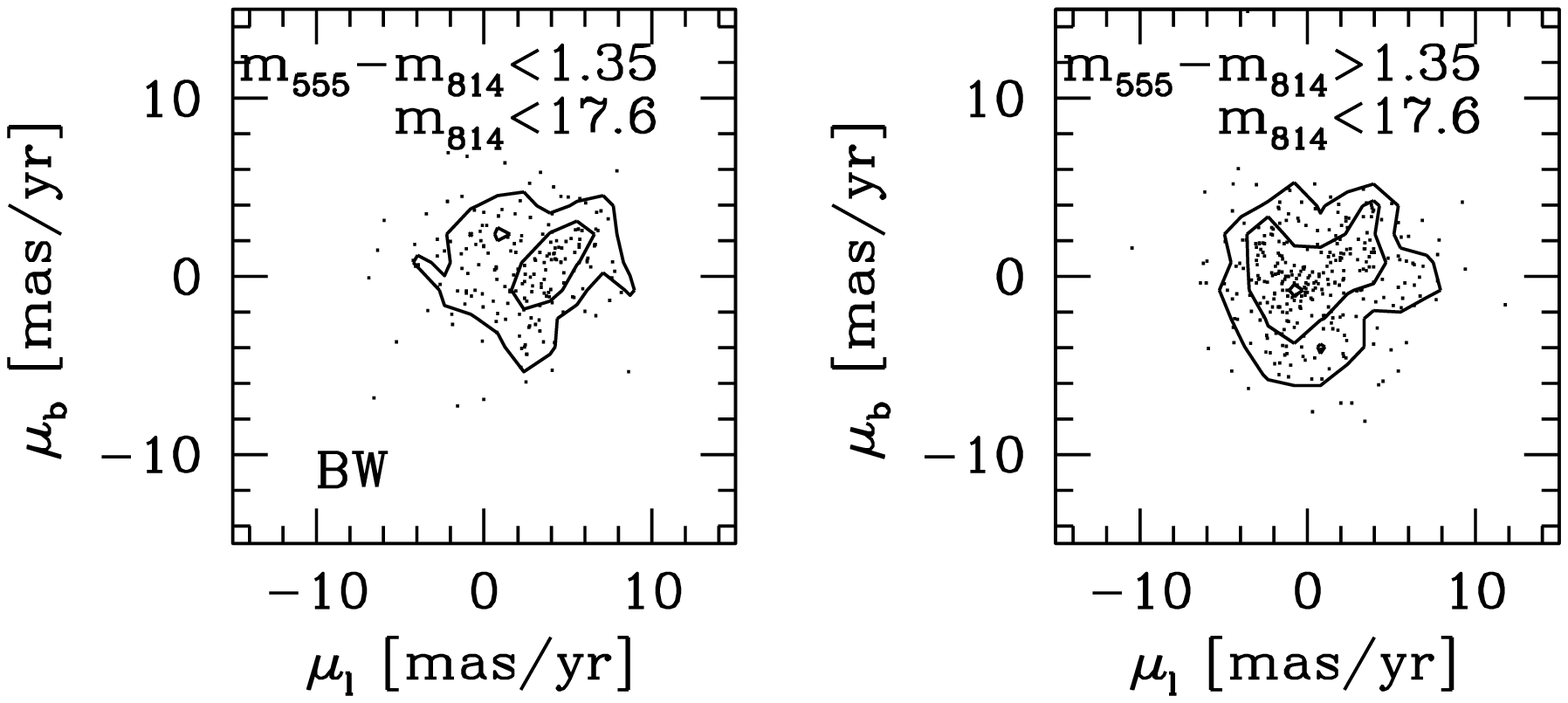}
\epsfxsize=\hsize\epsfbox{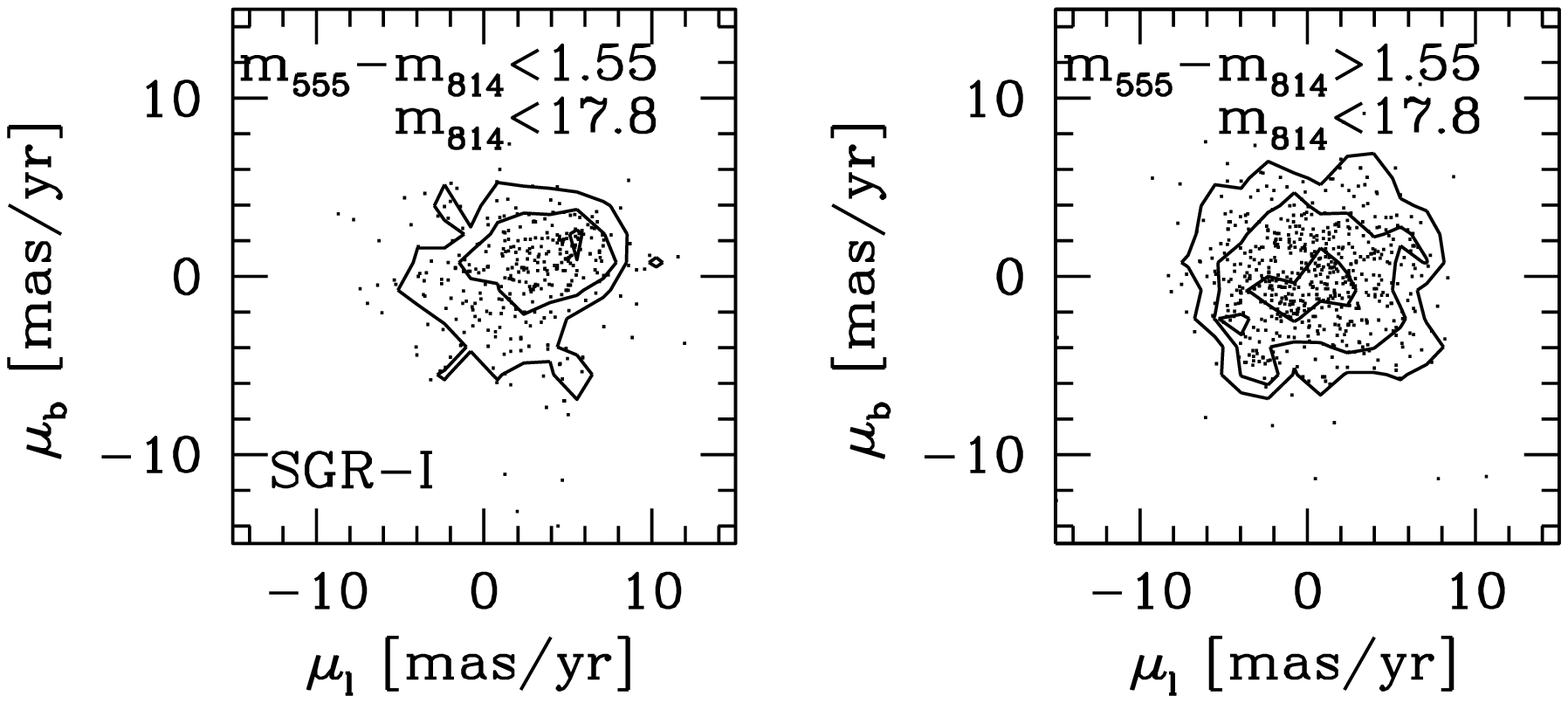}
\caption{The proper motion distributions of blue (left) and red
(right) giants in Baade's Window (top) and the Sgr-I field
(bottom). In both fields the blue giants show a marked proper motion
offset in the positive longitude direction, as expected for a
foreground population rotating in front of the bulge. The red giants
do not show this effect, suggesting they represent the dominant
population in these bulge windows.}
\end{figure}

\begin{table}
\begin{centering}
\caption{Proper motion dispersions (mas yr$^{-1}$) by field}
\label{tab:perccdstats}
\begin{tabular}{crrrrrr}
\tableline\tableline
 & \multicolumn{3}{c}{Baade Window}
 & \multicolumn{3}{c}{SGR-I Field}\\
Field & \multicolumn{1}{c}{$N$} & 
\multicolumn{1}{c}{$\sigma_l$} & \multicolumn{1}{c}{$\sigma_b$} & 
\multicolumn{1}{c}{$N$} & 
\multicolumn{1}{c}{$\sigma_l$} & \multicolumn{1}{c}{$\sigma_b$} \\
\tableline
PC  & 1076 & 2.91$\pm$0.06 & 2.51$\pm$0.05 & 1388 & 3.10$\pm$0.06 & 2.73$\pm$0.05 \\
WF2 & 5036 & 2.98$\pm$0.03 & 2.59$\pm$0.03 & 6583 & 3.25$\pm$0.03 & 2.78$\pm$0.03 \\
WF3 & 4848 & 3.00$\pm$0.03 & 2.53$\pm$0.03 & 6156 & 3.23$\pm$0.03 & 2.71$\pm$0.03 \\
WF4 & 4902 & 2.97$\pm$0.03 & 2.53$\pm$0.03 & 6107 & 3.27$\pm$0.03 & 2.84$\pm$0.03 \\
\tableline
All &15862 & 2.98$\pm$0.017 & 2.54$\pm$0.014 & 20234 & 3.24$\pm$0.016 & 2.77$\pm$0.014 \\
\tableline
\end{tabular}
\end{centering}

{Mean proper motions have not been measured. None of
the samples shows significant evidence for a misalignment of the
principal kinematic axes with galactic coordinates, i.e., we detect no
covariance between $\mu_l$ and $\mu_b$.}
\end{table}

Proper motion distributions and color-magnitude diagrams are
presented in Figures 2 and 3. The
velocity dispersions per CCD are listed in
table~\ref{tab:perccdstats}, and are consistent with each
other---there is no indication of systematic differences between the
velocity dispersions measured on the WF and PC CCD's for the same
fields, but there are field-to-field differences in the measured
velocity dispersions. The proper motion distributions are skewed, with
a longer tail towards positive $\mu_l$. We argue below that this is
in large part due to foreground contamination.

At first inspection, the color-magnitude diagram shows a turnoff
population with a blue extension and a red giant branch.  A first
cut at separating the kinematics of the various stellar populations
can be attempted using those stars brighter than the old turnoff
point, which we expect to show clear differences in kinematics
(Figure 4).  In fact, the blue main sequence stars
have a mean positive longitudinal proper motion of some 5mas/yr
with respect to the red giants. This is most simply understood as a
foreground, thin disk population which rotates in front of the bulge
stars.

\begin{figure}
\epsfxsize=\hsize\epsfbox{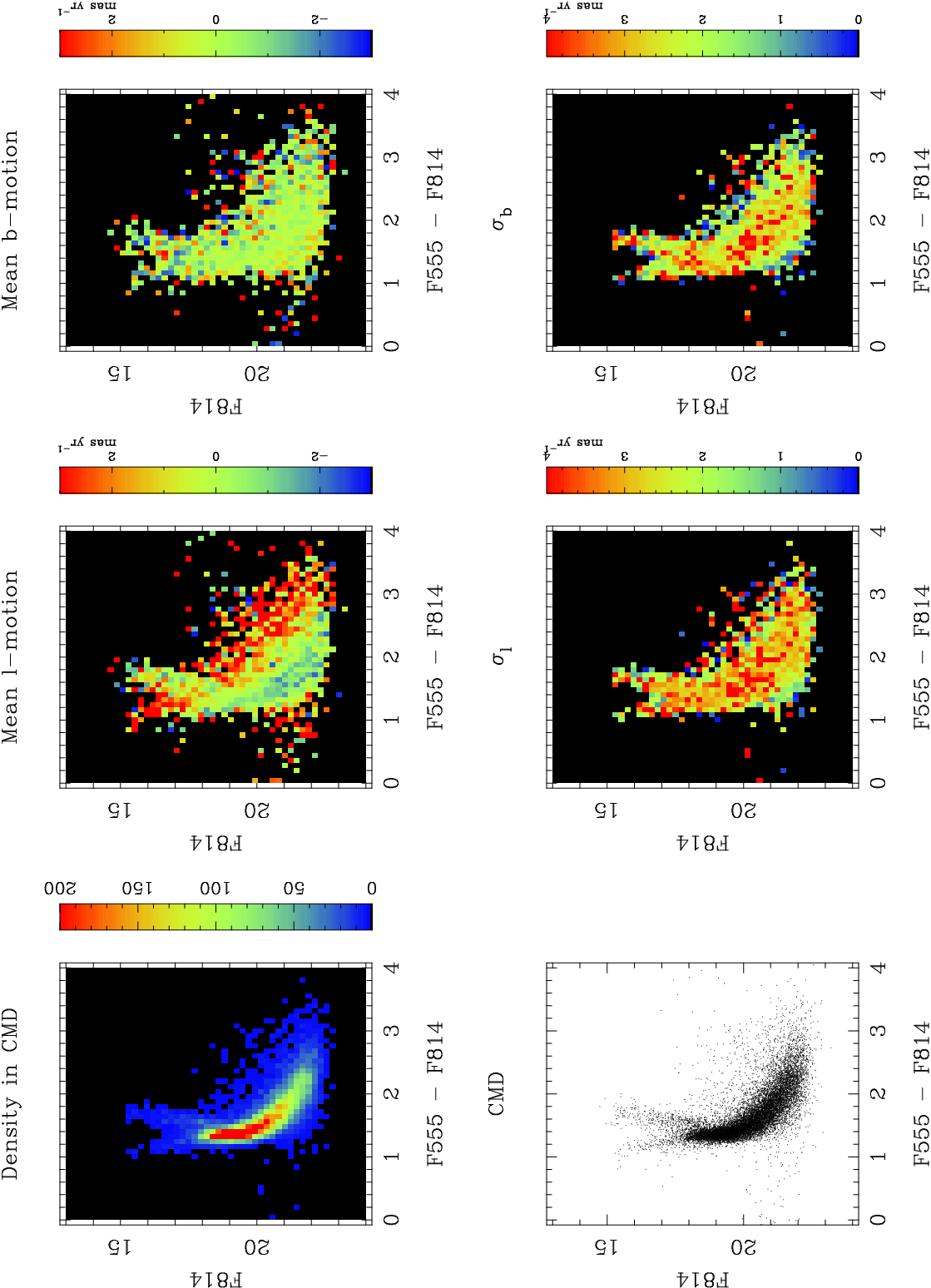}
\caption{Binned color-magnitude diagrams of Baade's Window. The
sample is plotted in each panel, color-coded in different ways. Top
row, left to right: number of stars in each bin; mean longitudinal
proper motion; mean latitudinal proper motion. Bottom row, left to
right: unbinned CMD; dispersion in longitudinal proper motion;
dispersion in latitudinal proper motion. Numbers above each plotted
point in the lower panels indicate the number of stars that went into
each $M^*$ bin.  Notice the clear separation of a disk main sequence
component in this field, and in Sgr I which is illustrated in Figure 6.}
\end{figure}

\begin{figure}
\epsfxsize=\hsize\epsfbox{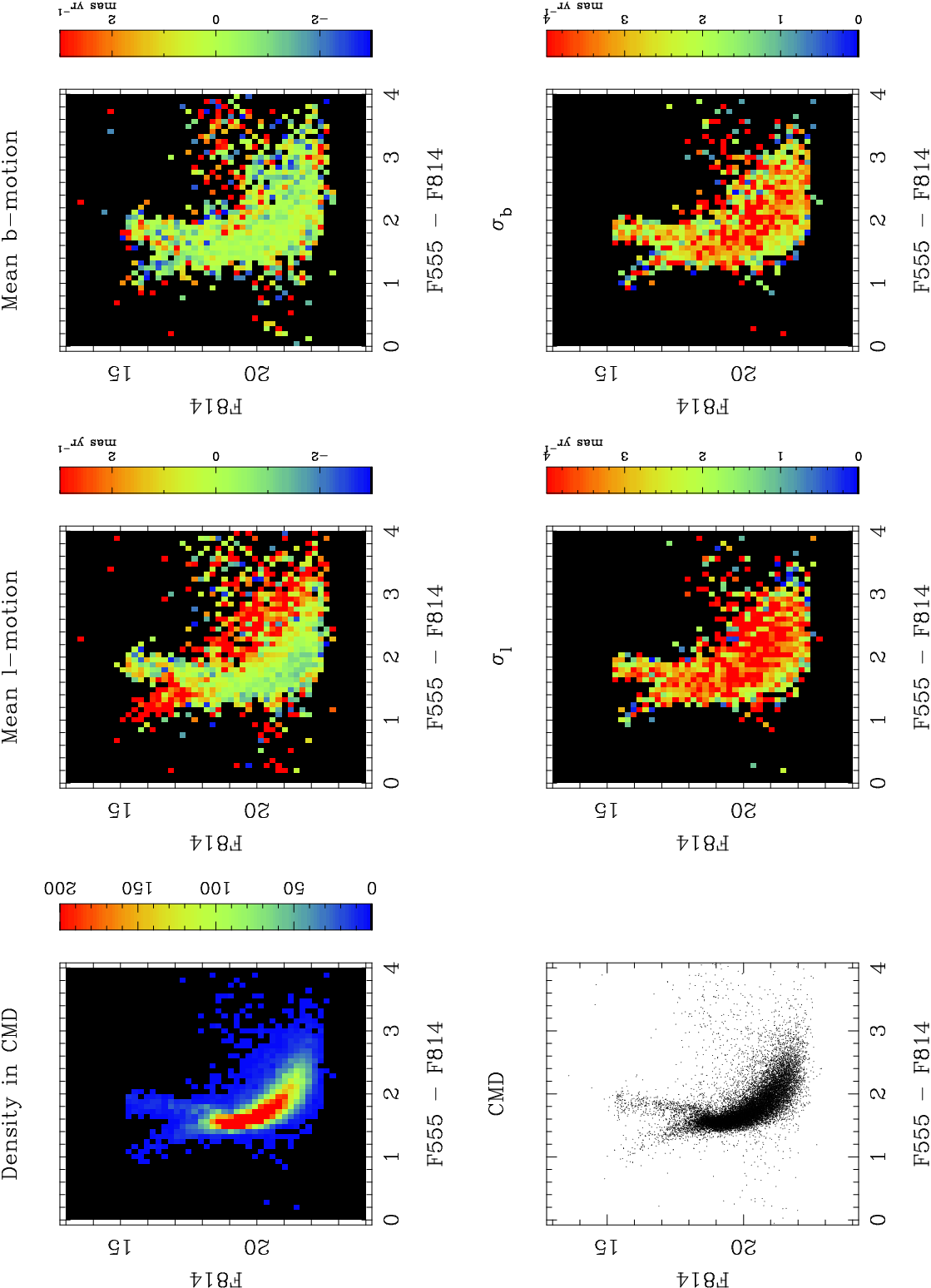}
\caption{Binned color-magnitude diagrams of the Sgr I field.
The sample is plotted in each panel, color-coded in different ways. Top
row, left to right: number of stars in each bin; mean longitudinal
proper motion; mean latitudinal proper motion. Bottom row, left to
right: unbinned CMD; dispersion in longitudinal proper motion;
dispersion in latitudinal proper motion. Numbers above each plotted
point in the lower panels indicate the number of stars that went into
each $M^*$ bin.}
\end{figure}

The bright blue main sequence stars are the part of the younger disk
population that stand out in the color-magnitude diagram (CMD), but
of course disk stars are found among fainter stars as well. This is
best illustrated in Figs. 5 and 6,
which are binned color-magnitude diagrams, with each bin color-coded
by various kinematic quantities. These figures clearly show the extent
of disk contamination across the CMD, as well as a number of other
kinematic features:
\begin{enumerate}

\item The disk kinematics of the bright blue main sequence stars
(large positive $\mu_l$) are shared by redder, fainter stars located
above the bulge main sequence. This is naturally explained as the main
sequence of disk stars in front of the bulge. As expected, the effect
is more pronounced in the Sgr-I field, which lies closer to the
galactic plane.

\item There is a gradient in the mean $\mu_l$ of faint main sequence
stars with magnitude: at a given color, the faintest main sequence stars
drift towards negative $l$. This suggests that we are measuring
stars clear through the bulge rotation field, even to the far
side of the bulge, where stars are rotating behind the galactic minor
axis.

\item The proper motion dispersions of the main sequence stars
decrease with magnitude: at a given color the faintest main sequence
stars have smaller dispersion in proper motion. We interpret this as a
distance effect: the faintest stars are further away and hence a given
velocity translates into a smaller proper motion.

\item There is a distinct suggestion in these diagrams of a
kinematically homogeneous bulge population which follows a single
isochrone, characteristic of an old stellar population.

\end{enumerate}

In order to investigate the kinematics further, we make cuts of main
sequence stars for which a crude distance modulus can be
calculated. We find that the quantity
\[
M^*=m_{F814W} - 2\left(m_{F814W}-m_{F555W}\right)
\]
removes the slope of the main sequence in the CMD, so that it may be
used as a simple relative distance indicator. As a function of this
distance modulus we plot the proper motions in
Figs. 7 and 8.  In both
Baade's Window and Sgr I,
the same picture emerges. We see a smooth gradient of rotation
down the line of sight. This is either due to contamination of a
non-rotating bulge by disk stars, or represents a true rotation of the
bulge. The notable similarity of the slope of both `rotation curves'
in both fields (even though the Sgr-I field is expected to be
more disk-dominated) argues for an intrinsic bulge rotation.

\begin{figure}
\epsfxsize=\hsize\epsfbox{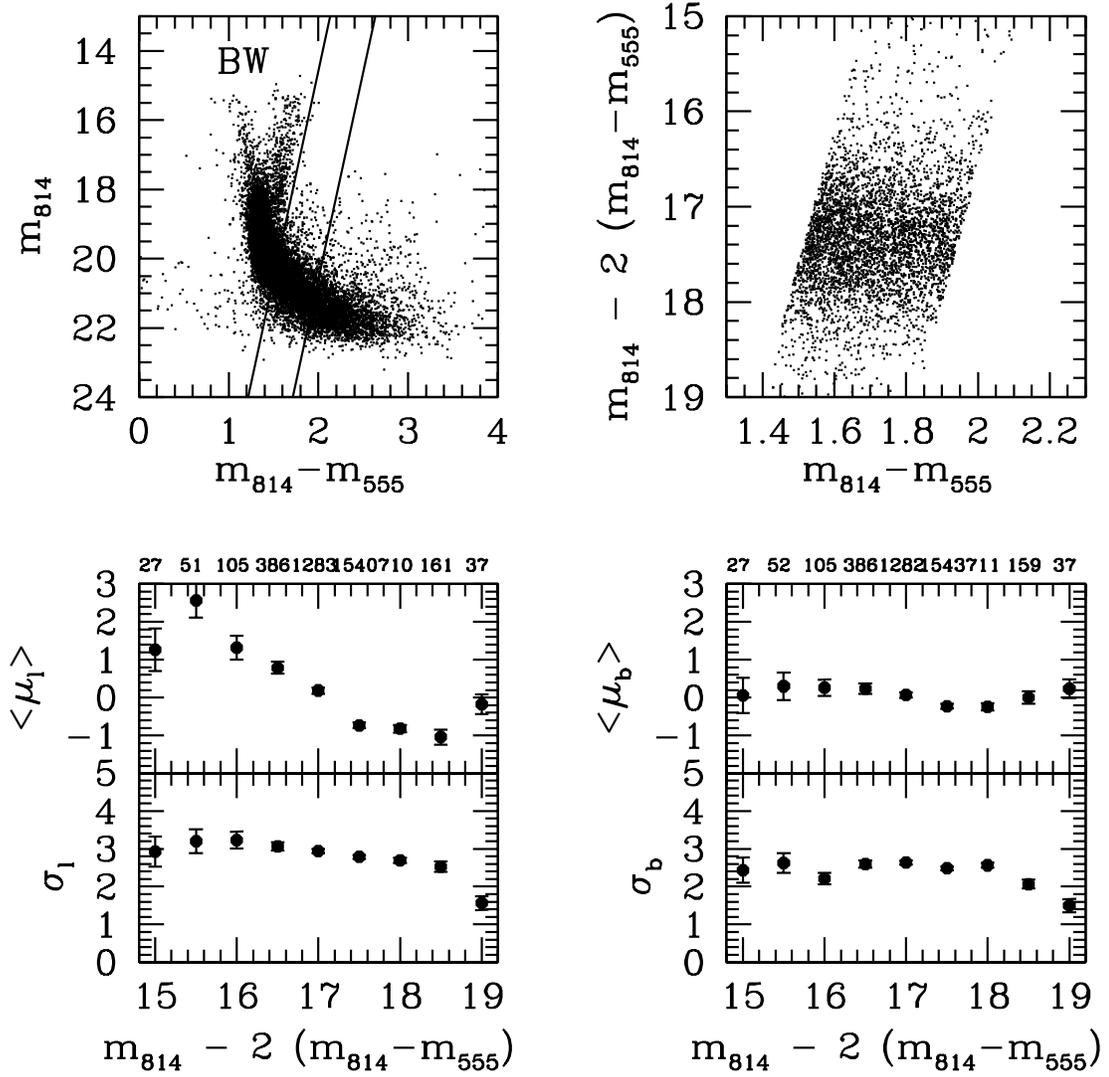}
\caption{The proper motion statistics in Baade's window versus the
distance estimator $M^*$. Top left: the CMD with the selection region
of stars. Top right: demonstration of the independence of $M^*$ with
color. Bottom left: Mean and dispersion in $\mu_l$ in `distance'
bins. Bottom right: Mean and dispersion in $\mu_b$ in 'distance' bins.}
\end{figure}

\begin{figure}
\epsfxsize=\hsize\epsfbox{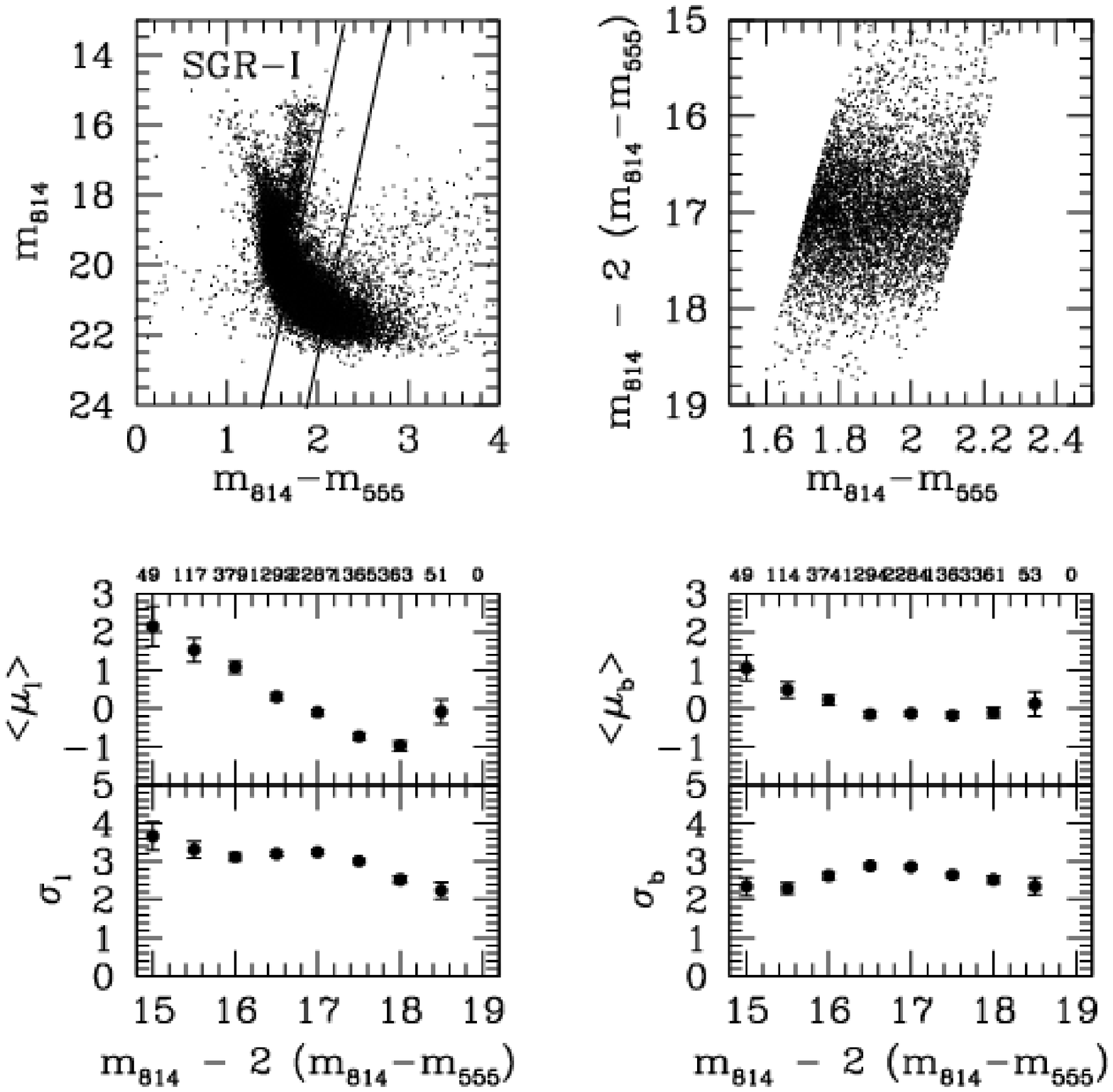}
\caption{As in Figure 7,  but for the Sgr-I field.}
\end{figure}

Radial velocity surveys find rotation in the bulge population; perhaps the
best example is that of \citet{izum95}, which shows the rotation
curve based on a survey of SiO masers.  The practicial difficulty with
measuring the rotation curve of the {\em pure bulge} population
is that as one moves off-axis, the density of the bulge drops rapidly
and sample contamination (from the halo and disk) becomes a major problem.
Our study gives one of the cleanest demonstrations of rotation in a bulge
star dominated sample.

Some undulations in the vertical mean proper motion are also seen.
There is the hint of a rise in the vertical proper motion for those
stars which might correspond to the ``middle'' of the bulge.  At this
point it is not clear whether these trends in the vertical mean proper
motion are real, or a manifestation of an unknown systematic residual
error.

It is notable that the proper motion dispersions decrease
for the faintest $M^*$. This is as expected if these stars are indeed
at larger distances: these stars would lie a factor two further away
than stars with $M^*$ ca. 1.5 magnitudes brighter.

Detailed modelling of these results is complicated by the uncertain
scatter in these crude photometric parallaxes.
However, we can estimate values for the velocity dispersions of bulge
stars in each field by considering the $M^*$ bin with zero mean
$\mu_l$. These stars should lie as close to the Galactic minor axis as
these lines of sight reach. The numbers are given in
Table~\ref{tab:veldisp}. We see evidence for a subtle anisotropy in
the bulge, but this could be an effect of the velocity gradient
through the $M^*$ bin. 

\begin{table}
\caption{Proper motion dispersions in km/sec (for $R_0=8$kpc)}
\label{tab:veldisp}
\begin{tabular}{lcccc}
\tableline\tableline
 & $\sigma(\mu_l)$ & $\sigma(\mu_b)$ & $\sigma_l$ & $\sigma_b$ \\
\tableline
Baade Window & 2.94 & 2.63 & 111 & 100\\
Sgr-I Field  & 3.24 & 2.85 & 123 & 108\\
\tableline
\end{tabular}
\end{table}

The tangential dispersions in Table~\ref{tab:veldisp} are similar to
those for the full sample (Table~\ref{tab:perccdstats}). This
coincidence is caused by two effects which act in opposite
directions. The full sample contains many distant stars, whose proper
motions are small and hence reduce the dispersion; on the other hand
the line-of-sight gradient in tangential velocity tends to increase
the width of the distribution. Only the first effect operates in the
vertical direction, so $\sigma(\mu_b)$ values are larger in
Table~\ref{tab:veldisp} than they are in the full sample.

\begin{figure}
\epsfxsize=0.45\hsize\epsfbox{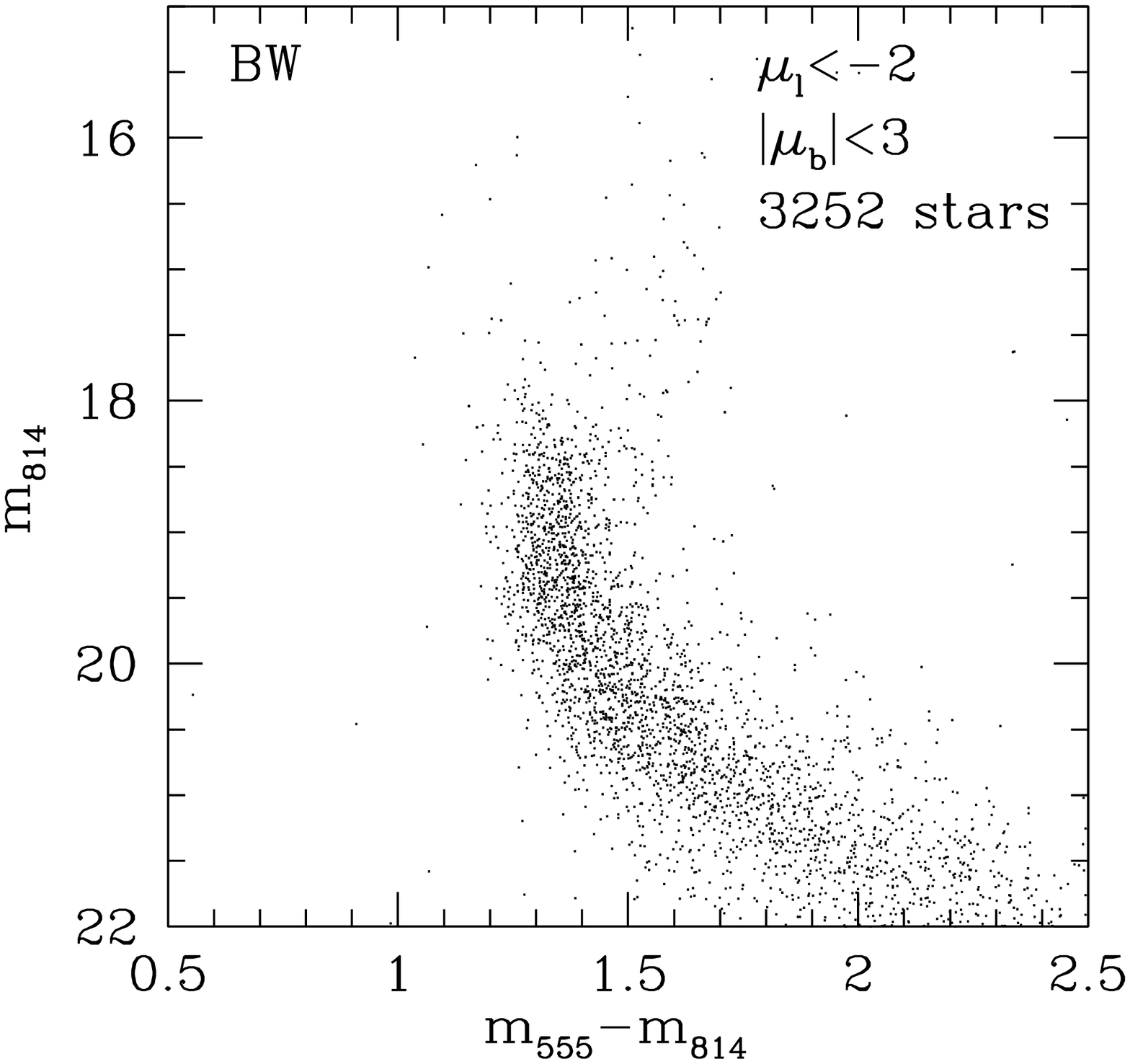}
\epsfxsize=0.45\hsize\epsfbox{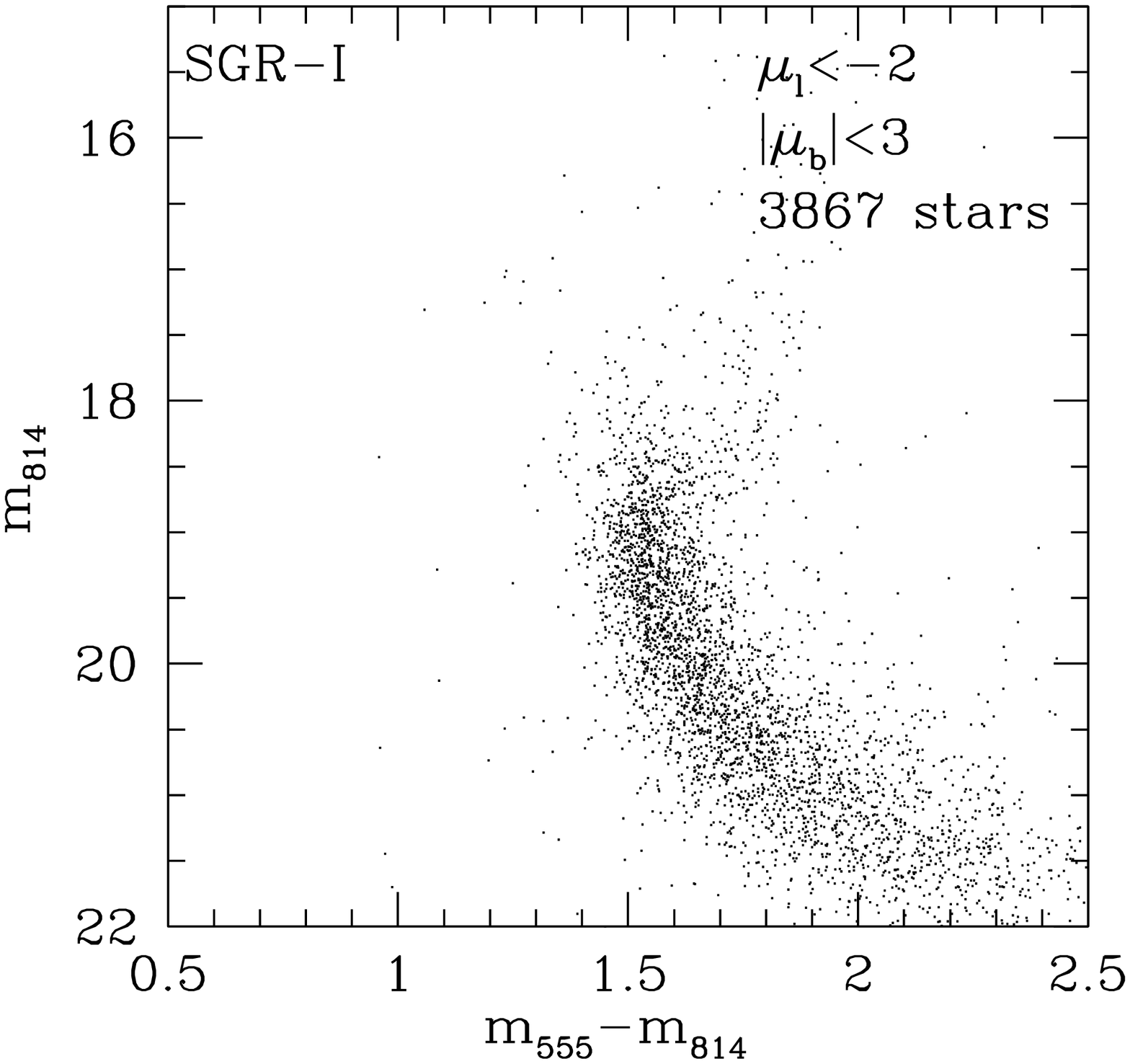}
\caption{The color-magnitude diagrams for kinematically-selected
bulge stars the two fields. The stars plotted are subsamples selected
kinematically against disk stars. The cut in $\mu_l$ removes
foreground stars with kinematics such as those of the blue giants,
while the $\mu_b$ upper limit cuts nearby M dwarfs seen particularly
in the Sgr-I field.}
\end{figure}

\section{The stellar population of the bulge}

With our picture of the kinematics in the bulge region, we now study
the stellar populations. We isolate a ''pure bulge'' sample by keeping
only stars with negative $\mu_l$ and moderate $|\mu_b|$. Given the
rotation gradient of the bulge, this kinematic cut may also
preferentially remove near-side bulge stars from the sample, so it may
be slightly biased in distance. Far-side disk contamination should be
very small as the line of sight rises away from the galactic plane
behind the bulge.

\begin{figure}
\epsfxsize=0.45\hsize\epsfbox{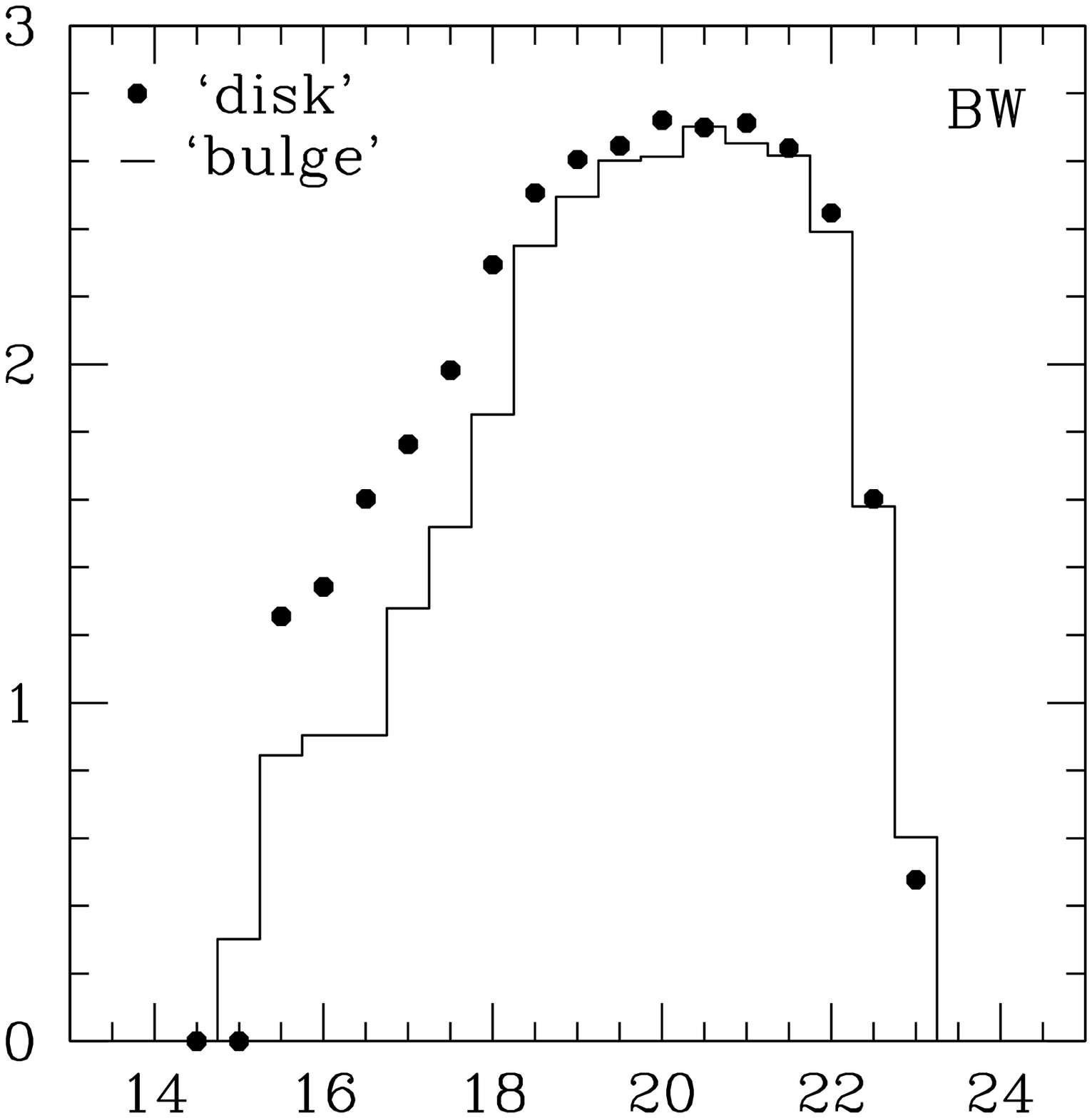}
\epsfxsize=0.45\hsize\epsfbox{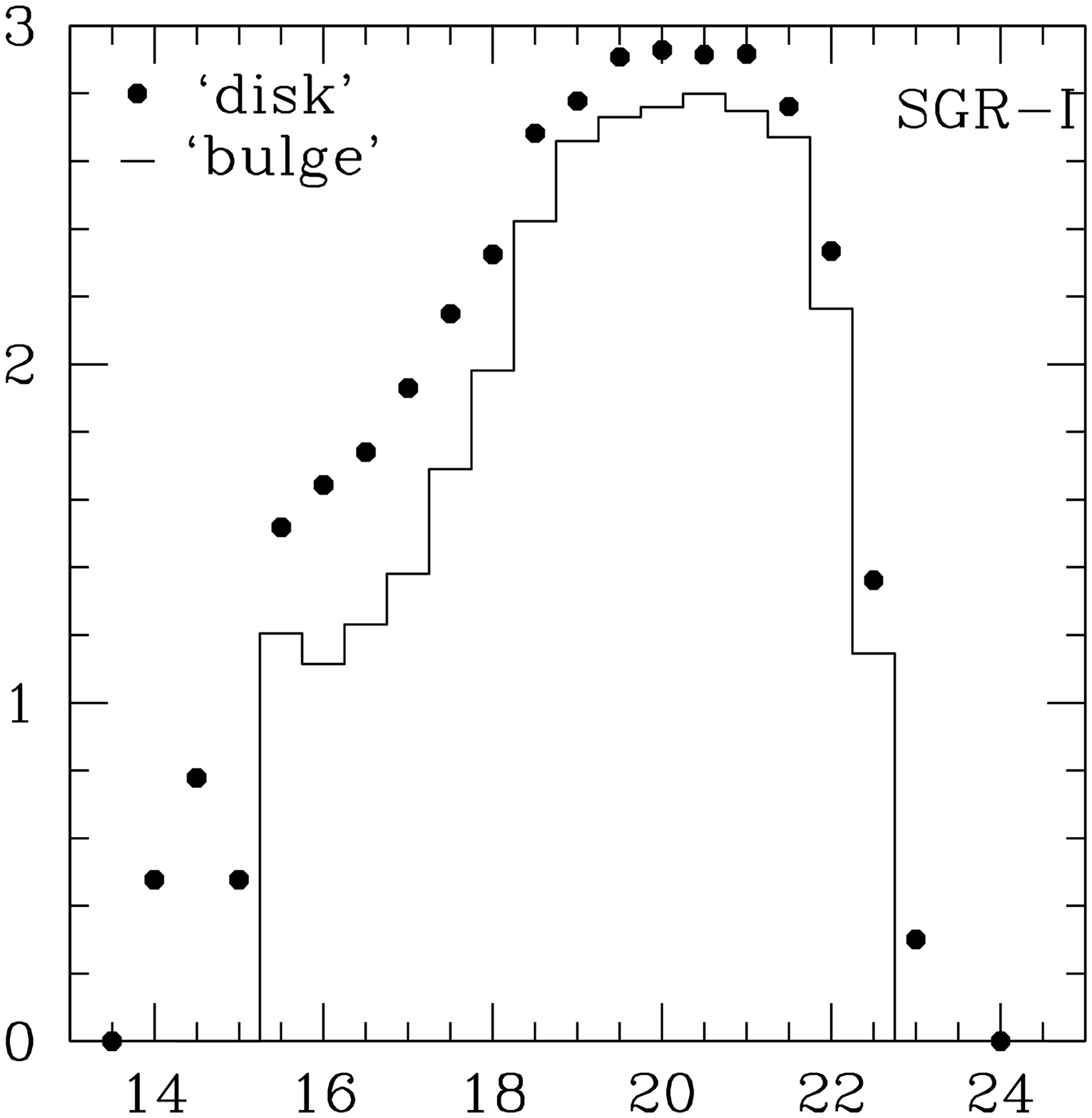}
\caption{Differential luminosity functions for stars selected
according to
the kinematic cuts used in Figure 9.
The shallower slope of the luminosity function at the
bright end is presumably due to an excess of young and intermediate
age stars in the disk, relative to the bulge (which rises steeply
at the old turnoff point).  We emphasize that the differences
are measured from stars in the same images and
are present at the same magnitude; therefore they cannot be due to image
crowding or incompleteness errors.
Notice the similarities between the
two fields, especially the steepness of the turnoff rise.}
\end{figure}

The color-magnitude diagrams that these cuts produce in
Baade's Window and the Sgr I field are shown in
Figure 9.
The $\mu_l, \mu_b$ cuts cleanly remove the bright blue portion
of the main sequence and foreground M dwarfs, by kinematic
selection only.  The color-magnitude diagram that remains strongly
resembles that of the old bulge globular cluster NGC 6553
(cf. \citealt{zocc01}).  A handful of stars brighter than the
old turnoff remain (field blue stragglers or outliers?) but the
dominant blue extension is clearly demonstrated to be a foreground
population.   The present color-magnitude diagram suggests that
blue stragglers are not common in the field population (for example,
not nearly as frequent as found in the cluster NGC 6553 by \citealt{zocc01}).
However, it does not put interesting limits on their numbers.  To do so
would require radial velocities and perhaps line strength measurements.
When additional analysis is finished, we will be able to derive the
age of the bulge field relative to NGC 6553 and 6528; a cursory
examination of  Figure 9 shows the characteristic turnoff point
of a clearly old stellar
population.

The impression that two very different populations are being
identified via the proper motions is amplified when one considers
the luminosity functions for the bulge and disk populations.
Recall that (especially at the bright end) incompleteness and
photometric errors are identical, being only magnitude
dependent.  Figure 10 shows our plot for the
Baade's Window disk and bulge samples, and the disk's extension
(relative to the sharp rise of the main sequence turnoff) is
evident.  These plots give further strength to
the visual impression from Figure 9.

We emphasize that our population separation is based on kinematics,
not on any photometric or spectroscopic classification. While disk and
bulge exist cospatially, they separate well kinematically, and this is
therefore at least a complementary, and we would argue a more robust,
technique for isolating a `pure bulge' population.

\section{Conclusions}

With HST photometry and proper motions determined with high enough
precision, it is possible to separate the disk and bulge populations
by their kinematics alone.  The long standing question regarding the
nature of the blue main sequence extension in the bulge field
population is settled: the great majority of those stars evidently
belong to the foreground disk.  When these stars are excluded, the old
turnoff population in the bulge remains, and no measureable population
of blue stragglers or intermediate age stars is present.
This was first demonstrated for the Baade's Window field by
\cite{ortolani95}.
\cite{feltzgil00} came to the same conclusion based on
counts of stars brighter and fainter than the turnoff point in their
WFPC2 data. However, a proper accounting for the foreground
disk has been a persistant issue,
and our application of the kinematic data strengthens
greatly the conclusion that the bulge is dominated by old stars.

We take our study a step further, and find direct evidence for the
rotation of the bulge population. The observed proper motion
anisotropy of bulge stars is largely caused by the line-of-sight
gradient of the rotation of the bulge; when this is removed a nearly
isotropic velocity distribution of the bulge stars results. The
velocity dispersion declines from Sgr~I to Baade's Window.

When appropriate samples in Baade's Window are compared, our proper
motion dispersions agree with those found by \citet{spaen92} and by
\cite{feltz01}.  Our results compare well with predictions of the
\citet{zhao96} and \citet{zhaoetal96} bulge model, but since our
proper motion samples are more than an order of magnitude larger than
existing radial-velocity and proper-motion samples of bulge stars, and
extend well below the turnoff, the time is ripe for more involved
modelling. Extension of this work to further bulge fields
(e.g. \citealt{zocc01}), and combination of these results with
spectroscopy (for radial velocities, metallicity, and improved
distance estimates) should open the way for a new chapter in our study
of the Galactic bulge.

\acknowledgements

Support to RMR for proposal GO-8250 was provided by NASA through a
grant form the Space Telescope Science Institute, which is operated by
the Association of Universities for Research in Astronomy, Inc., under
NASA contract NAS 5-26555.

\appendix

\newcommand{\PSF}{P}

\section{Accuracy of PSF-fitting photometry and positions}

Here we calculate the accuracy of object centroids derived by means of
PSF fitting.

Let $f_i$ be the data: intensities on the pixels $i$ at positions
$x_i,y_i$ on the image plane. We try to model these data as
$ A \PSF(x_i-\mu_x,y_i-\mu_y) \Delta^2$, where $\Delta$ is the pixel
width, and $\PSF$ is the PSF normalized to total intensity 1.
$A$ is the intensity of the star, and $(\mu_x,\mu_y)$ are the position
of the star, to be fitted for.
Write
\[
\chi^2=\sum_i \left(f_i - A \PSF(x_i-\mu_x,y_i-\mu_y) \Delta^2\right)^2
                /\sigma_i^2
\]
where $\sigma_i$ is the error on the measured intensity in pixel $i$.
Then the minimum of $\chi^2$ gives the best-fit PSF, and the second
partial derivatives of $\chi^2$ with respect to any two parameters
give twice the inverse covariance matrix. In what follows we ignore
covariances between the different variables $A$, $\mu_x$ and $\mu_y$,
as is appropriate for bisymmetric PSF's.

At the best-fit value (we may assume without loss of generality that
$\mu_x=\mu_y=0$) we obtain the inverse variance on the star's intensity
$A$ as
\[
\hbox{Var}(A)^{-1} =
0.5 \partial^2\chi^2/\partial A^2=\sum_i \PSF (x_i,y_i)^2 \Delta^4 /\sigma_i^2
\]
which reduces to
\[
                 =\left(\int \PSF^2 dx dy\right) \Delta^2 / \sigma^2
\]
if the PSF is fully sampled and $\sigma_i$ constant (i.e.,
background- or read noise-limited data).

The inverse variance of the best-fit position is similarly (removing
terms which go to zero at the best fit)

\[
\hbox{Var}(\mu_x)^{-1} =
0.5 \partial^2\chi^2/\partial\mu_x^2 =
\sum_i A^2 (\partial\PSF/\partial x)^2 \Delta^4 /\sigma^2
  =\left(\int (\partial\PSF/\partial x)^2 dx dy\right) A^2 \Delta^2 / \sigma^2
\]
(and similarly for $\mu_y$).
These two relations can be combined to give
\[
         \delta \mu_x={\delta A\over A} \times \sqrt{\int \PSF^2 dx dy \over
\int (\partial\PSF/\partial x)^2dx dy}
\]
where $\delta A = \hbox{Var}(A)^{-1/2}$ is the 1-$\sigma$ error on
$A$, etc.

The centroid error therefore depends on the significance $A/\delta A$
of the detection of the star, and on a geometric factor governed only
by the shape of the PSF.

For a gaussian PSF, dispersion $s$, we find
\[
\hbox{Var}(A)^{-1}=\Delta^2/(4 \pi s^2 \sigma^2)
\]
hence
\[
\delta A    =\sqrt{4 \pi} s \sigma /\Delta
\]
and
\[
\hbox{Var}(\mu_x)^{-1}=A^2 \Delta^2 /(8 \pi s^4 \sigma^2)
\]
hence
\[
\delta \mu_x    =\sqrt{8 \pi} s^2 \sigma / (\Delta A)
                    =\sqrt2 s {\delta A\over A}
\]

For a Moffat function PSF of the form
\[
\PSF={\beta-1\over \pi a^2} \left(1+{r^2\over a^2}\right)^{-\beta}
\]
we get
\[
      \delta \mu_x=a {\delta A\over A} \sqrt{2 \beta+1\over \beta (2\beta-1)}
\]
If we write $a$ as $\frac12 \hbox{FWHM}/\sqrt{2^{1/\beta}-1}$,
we find that for $\beta> 1.5$, which covers PSF's with tails as
shallow as $r^{-3}$
\[
         \delta \mu_x={\delta A\over A} \times 0.67 \times \hbox{FWHM} \pm 10\%.
\]
(The gaussian is the limit $\beta\to\infty$; in this case the
coefficient is 0.6.)

\end{document}